\documentstyle[pramana,epsf,floats]{ias}
\begin{document}
\mark{{String Inflation}{String Inflation}}
\title{Inflatable String Theory?}

\author{C.P. Burgess}
\address{Physics Department, McGill University,\\ 3600 University Street,
Montr\'eal, Qu\'ebec, Canada, H3A 2T8.}
 \pacs{}
\abstract{The inflationary paradigm provides a robust description
of the peculiar initial conditions which are required for the
success of the Hot Big Bang model of cosmology, as well as of the
recent precision measurements of temperature fluctuations within
the cosmic microwave background. Furthermore, the success of this
description indicates that inflation is likely to be associated
with physics at energies considerably higher than the weak scale,
for which string theory is arguably our most promising candidate.
These observations strongly motivate a detailed search for
inflation within string theory, although it has (so far) proven to
be a hunt for a fairly elusive quarry. This article summarizes
some of the recent efforts along these lines, and draws some
speculative conclusions as to what the difficulty finding
inflation might mean.}

\maketitle \vspace{-0.5cm}\rightline{McGill-04/19}
\section{Introduction}

Recent years have seen many detailed comparisons between
observations and the Hot Big Bang model of cosmology, within which
the early universe consists of a hot soup of elementary particles.
This picture bears up to scrutiny very well, at least for all
times after the epoch of Big Bang Nucleosynthesis (BBN). Embedded
within this success are some baffling puzzles, however, including
the following three:
\begin{itemize}
\item What is the nature of the Dark Energy whose energy density
appears to presently make up over 70\% of the energy density of
the universe?

\item What is the origin and nature of the Dark Matter which
dominates the mass of galaxies and which makes up the lion's share
of the universal energy density once the Dark Energy is removed?

\item Why does the universe start out in the extremely homogeneous
and isotropic initial state which observations require?

\end{itemize}

The Inflationary Universe proposal \cite{inflation} addresses
itself to the third of these questions. It is based on the
compelling observation that the flatness, isotropy and homogeneity
of the universe during the current epoch could naturally and
robustly follow if the universe were to undergo a dramatic,
accelerated expansion during some earlier epoch. Furthermore, an
accelerated exponential expansion can actually occur for
reasonable choices for the equation of state for the universe's
matter content. In particular, it arises if the potential energy
of some scalar field were to temporarily dominate the energy
density of that part of the universe within which we now live.

This much was known more than 20 years ago, but the relatively new
development is the realization that the same inflation which was
designed to solve the horizon and flatness problems can also very
naturally explain the observed fluctuations in the cosmic
microwave background (CMB). It can do so because the required
dramatic universal expansion required by inflation can also
stretch ordinary quantum fluctuations from microscopic up to
cosmological length scales. Indeed, this picture can provide an
attractive explanation for many features of the {\it spectrum} of
these fluctuations, including their nearly scale-invariant
spectrum, their Gaussianity, and so on \cite{cmbinflation}.

The successful description of the {\it amplitude} of inflationary
fluctuations is related to the value of the universal expansion
rate, $H = \dot a/a = [V/(3 M_p^2)]^{1/2}$, at that epoch during
inflation when the currently-observed CMB fluctuations were
leaving the horizon.\footnote{In what follows $M_p$ represents the
rationalized Planck mass, $(8\pi G)^{-1/2} \approx 10^{18}$ GeV.}
As such, it is also sensitive to the energy scale, $V$, of the
physics whose equation of state generates the inflationary
behavior. Typically, for inflation associated with the slow roll
of a scalar field, one finds that the combination $\delta^2 =
(1/150\pi^2) (V/M_p^4 \epsilon)$ must satisfy $\delta \approx 2
\times 10^{-5}$ in order to reproduce correctly the observed
fluctuation amplitude \cite{L&L}. The small quantity $\epsilon =
\frac12 (M_p V'/V)^2$ which appears here is a measure of how flat
the potential $V$ is as a function of the relevant
(canonically-normalized) slowly-rolling scalar field, or inflaton.

For $\epsilon$ not too small one finds in this way $V^{1/4}$ only
a few orders of magnitude smaller than $M_p$, which places it well
beyond the reach of relatively well-understood physics. Indeed,
being so close to the Planck scale, this suggests the physics
appropriate to inflation is related to the physics of gravity at
high energies. Happily, there are not many candidates for
consistently describing gravitational physics at such high
energies, with only one theory so far --- string theory --- having
been explored well enough to try to make detailed contact with
inflationary dynamics.

The possibility of such a connection is extremely exciting.
Whereas inflation can be described as a successful
phenomenological explanation for observations which is groping for
a theoretical framework, string theory is an extremely
well-motivated theory of very short-distance physics searching for
observations which it can explain. The possibility that inflation
might provide the long-sought-for bridge between string theory and
observations provides a strong incentive for exploring the
circumstances under which string theory might give rise to
inflation.

As might be expected given such good motivation, the search for
inflation within string theory is an old one
\cite{stringinflationreviews}. The main lesson of these early
searches is that inflation would not be easy to find, and in
retrospect this can be attributed to two basic reasons. First and
foremost of these was the problem of moduli: for technical reasons
the best-explored string vacua are supersymmetric, and these vacua
typically have many massless scalar fields with strictly vanishing
scalar potentials. Besides the quite general cosmological problems
which the existence of such moduli potentially raises
\cite{moduliproblems}, they also obstruct the search for
inflation. They do so because these scalars cannot remain massless
once supersymmetry breaks, and because the inflaton potential is
(by design) very shallow its slope can be dominated by relatively
small contributions to the scalar potential, such as can occur
when the moduli acquire their potential through the breaking of
supersymmetry. Furthermore, very general results imply that this
supersymmetry breaking cannot occur at all within perturbation
theory. As a result progress becomes difficult because an
understanding of the cosmology of these fields was held hostage to
the development of a non-perturbative understanding of how
supersymmetry breaks.

The second obstacle to model building in these early years was the
belief at the time that the string scale must be near the Planck
scale. This made it more difficult to achieve potentials for which
the fluctuation amplitude had the right size, because the natural
scale of the problem was $M_p$ and the small dimensionless
parameters were typically extremely small --- often $\sim
e^{-(4\pi)^2/g^2}$ for some small coupling constant $g$ --- given
the nonperturbative nature of the potentials required.

The search for stringy inflation has had a recent revival because
of developments within string theory itself, which have suggested
new ways around both of the above difficulties. The most important
development was the discovery of D-branes and other kinds of
branes. This discovery of branes --- which consist of surfaces to
which strings can attach --- fundamentally changes what the
low-energy limit of string theory can look like.

First, branes introduce nonperturbative string physics which is
much more accessible to calculation than were previous
nonperturbative approaches. In particular, some quantities like
brane tensions vary as inverse powers of string coupling and so
branes themselves are able to evade previous perturbative no-go
theorems which precluded supersymmetry breaking. Consequently the
examination of brane dynamics opens up the possibility for the
semiclassical study of supersymmetry-breaking effects like the
generation of potentials for moduli. For inflationary purposes the
most actively pursued direction along these lines has been to have
supersymmetry break due to the presence of both branes and
anti-branes, with the inflaton being associated with the relative
motion of the brane/anti-brane pair.

Second, the existence of branes opens up the possibility that the
string scale, $M_s$, can be much smaller than $M_p$. One way this
could happen would be through the {\it brane-world} scenario,
within which all known low-energy particles but the graviton were
to be trapped on the surface of a brane. In such a situation
different interactions (like electromagnetism and gravity) can
`see' different numbers of dimensions and this complicates the
inference of the string scale from the observed value of Newton's
constant, in such a way as to allow $M_s \ll M_p$. This allows the
possibility that scalar potentials of order $V \sim M_s^4$ could
lie in a range of interest for describing the amplitude of density
fluctuations within the CMB.

It is also clear that cosmology must change dramatically within
any brane-world picture, since appearance on the branes of
localized sources of energy implies the overall distribution of
energy throughout the universe is dramatically different from the
homogeneous and isotropic configurations which were entertained in
the earlier proposals. All of these considerations have spurred
new progress in understanding how inflation might arise within a
string-theoretic context.

The remainder of this paper is organized in the following way. The
next section, \S2, provides a very short description of how brane
physics has been used to make progress on the moduli-fixing
problem within the context of the compactification to 4 dimensions
of Type IIB string vacua. \S3 follows this with a very recent
simple proposal for using these ideas to obtain inflation, using
one of the geometrical moduli as the inflaton. \S4 describes the
better-explored alternative wherein inflation uses the relative
position of various branes as the putative inflaton. Some
preliminary conclusions are drawn from these explorations in the
final section, \S5. For reasons of space and focus the topics
described in this summary are limited to work on which I have been
directly involved myself, forcing the omission of other
interesting lines of development such as those of
ref.~\cite{omissions}.

\section{Modulus Fixing}

Although it was early realized that branes could break
supersymmetry and so potentially help address the modulus problem,
the decisive step forward required the embedding of this
observation into concrete compactifications of string vacua to 4
dimensions in a way which preserves $N=1$ supersymmetry. This was
achieved for Type IIB vacua \cite{GKP,Sethi1} by generalizing the
construction of the earlier Calabi-Yau heterotic compactifications
\cite{calabiyau} to include various D3 and D7 branes, together
with their orientifolds. The inclusion of these branes, as well as
the warping and various fluxes which they source, turn out to
generate a potential for the many complex-structure moduli of the
Calabi-Yau vacua, while leaving 4D $N=1$ supersymmetry unbroken
and without generating a potential for the various K\"ahler moduli
(of which there is always at least one).

Within this context further progress was then made by focussing on
those Calabi-Yau spaces having the fewest unlifted K\"ahler
moduli. In ref.~\cite{KKLT} KKLT focussed their attention on those
spaces having just one such modulus, $T$, with attention then paid
to how this last modulus might be lifted. This was studied in the
low-energy 4D supergravity described by these vacua using the
standard $N=1$ supergravity potential \cite{cremmeretal},
\begin{equation} \label{Ftermpot}
    V_F = e^K \, \Bigl[ K^{i\overline{j}} (W_i + K_i W) (W_j + K_j
    W)^* - 3 |W|^2 \Bigr] \,,
\end{equation}
where $K(\phi^i,\phi^{i*})$ is the low-energy supergravity's
K\"ahler potential and $W(\phi^i)$ is its superpotential. In this
expression the quantity $K^{i \overline{j}}$ represents the matrix
inverse of the K\"ahler metric, $K_{i\overline{j}}$, where here
and above subscripts denote differentiation with respect to
$\phi^i$ or $\phi^{i*}$.

The explicit form for $K$ and $W$ can be obtained semiclassically,
along the same lines as originally performed for Calabi-Yau
compactifications \cite{cyreduction}, with the result
\begin{equation} \label{GKPWK}
    W = W_0 \qquad \hbox{and} \qquad K = - 3 \ln (T + T^*) \,,
\end{equation}
where $W_0$ is $T$-independent. This has the form of a {\it
no-scale} model \cite{noscale}, for which the classical potential,
$V_F$, vanishes identically even though supersymmetry breaks for
any finite $T$. This flatness of the potential for $T$ reflects
the fact that the 4D compactifications of ref.~\cite{GKP} do not
lift K\"ahler moduli such as $T$.

The lifting of the degeneracy along the $T$ direction requires
additional physics, and in ref.~\cite{KKLT} this was assumed to be
achieved through nonperturbative effects, such as through gaugino
condensation \cite{gc,ourgc} of a low-energy gauge sector
localized on one of the D7 branes. In this case the superpotential
becomes modified from eq.~(\ref{GKPWK}) to
\begin{equation} \label{KKLTW}
    W = W_0 + A \, e^{-a T} \,,
\end{equation}
where $A$ and $a$ are calculable. This superpotential generates a
nontrivial potential, $V_F$, for $T$, typically stabilizing it at
finite $T$ with $V_F < 0$. The resulting minimum preserves $N=1$
supersymmetry and has an anti-de Sitter 4D spacetime geometry with
no remaining moduli.

The remaining supersymmetry may also be broken, with $V_F$ at the
minimum raised to positive values, through the artifice of
introducing an anti-D3 brane into the underlying string
configuration. (Generalizations of these results to more
complicated superpotentials have also been considered
\cite{saltatory}.) Alternatively, the same effect as the anti-D3
branes can be obtained by turning on magnetic fluxes on the D7
branes \cite{bkq}, or by choosing more complicated local minima of
the flux-induced potential \cite{Eva}.

As has recently been noticed \cite{Douglas,Sethi2}, it can happen
that the nonperturbative physics required to generate
superpotentials of the form of eq.~(\ref{KKLTW}) may not be
permitted for the simplest Type IIB vacua, and in particular might
not arise for the single-modulus vacua considered in
ref.~\cite{KKLT}.\footnote{According to ref.~\cite{Douglas} the
required nonperturbative physics can arise for two-modulus models,
however.} As of this writing it remains open whether this
conclusion can be avoided for more complicated constructions (such
as those involving orbifolding of the simplest ones) or if the
minimal constructions must require two or more moduli.

\section{Racetrack Inflation}

Perhaps the simplest approach to obtaining inflation within string
theory is to try to do so using the superpotential of
eq.~(\ref{KKLTW}), but this turns out not to lead to sufficiently
flat potentials. Slow-roll inflation does become possible for a
minor generalization of this superpotential, however
\cite{racetrackinflation}. The required generalization is what
would arise if the nonperturbative physics which is responsible
for the superpotential involves gaugino condensation for a
low-energy sector involving a product of gauge groups
\cite{gcproduct}. This leads to a superpotential which has the
modified racetrack form:\footnote{This is a `modified' racetrack
because the original racetrack models \cite{racetrack} -- which
did not inflate -- did not include the crucial parameter $W_0$.}
\begin{equation} \label{modracetrack}
    W = W_0 + A \, e^{-aT} + B \, e^{-bT} \,,
\end{equation}
where $A$, $B$, $a$ and $b$ are known constants ({\it e.g.} $a = 2
\pi/N$ and $b = 2 \pi/M$ for a pure gauge theory with gauge group
$SU(N) \times SU(M)$).

\begin{figure}[htbp]
\epsfxsize=8cm \centerline{\epsfbox{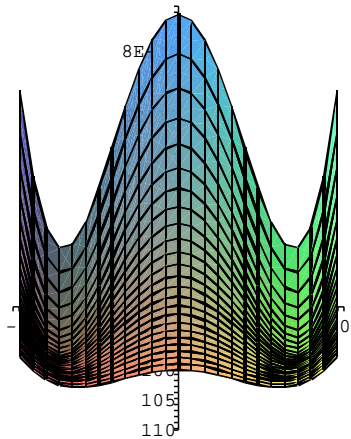}}
\caption{Plot for a racetrack type potential illustrating the
local minima between which a saddle point exists on which
inflation can start. Units are $M_p=1$.} \label{F1}
\end{figure}

The potential $V_F$ produced by this superpotential has stationary
points which can occur at large values of Re~$T$, provided we
choose $|N-M| \ll N,M$. This is a good thing, since a large value
for Re~$T$ is required for the consistency of the supergravity
analysis in terms of effective 4D fields because Re~$T$ is
proportional to the volume of the internal 6D space, and this was
the original motivation for developing racetrack models in the
first place \cite{racetrack}. Since the value of the potential at
the minimum is negative, this corresponds to an anti-de Sitter
vacuum in 4 dimensions.

\begin{figure}[htbp]
\epsfxsize=8cm \centerline{\epsfbox{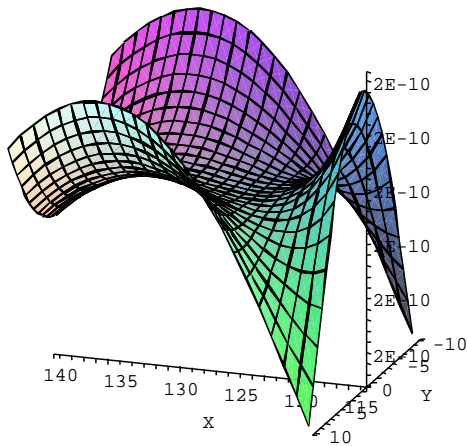}}
\caption{Plot for the same potential as for Fig.~\ref{F1}, showing
a close-up of the saddle point which lies between the two local
minima. Units are $M_p=1$.} \label{F2}
\end{figure}

It turns out that this minimum can be lifted to zero or to
positive values, much like the situation for the KKLT solution, by
adding an antibrane. For an appropriately chosen brane this adds a
new term $\delta V = E / (\hbox{Re}~X)^{2}$, where $E$ is
$T$-independent. Once this is done, slow-roll inflation becomes
possible starting from the saddle point which appears for some
choices of parameters between the various minima
\cite{racetrackinflation}. The minima are illustrated in
Fig.~\ref{F1}, while Fig.~\ref{F2} shows a closeup of the saddle
point in between. These figures are drawn using the parameter
values
\begin{equation}
\label{parvalues} A=\frac{1}{50}, \qquad B=-\, \frac{35}{1000},
\qquad a=\frac{2\pi}{100}, \qquad b=\frac{2\pi}{90},
  \qquad W_0 = -\frac{1}{25000}\   \ .
\end{equation}

It is possible to adjust the quantities $A$, $B$, $a$, $b$ and
$W_0$ so that the slow-roll parameters are small at the position
of the saddle point. Writing $T = X + iY$ and using canonically
normalized fields, the $\eta$ parameter at the saddle point is
given by the expression $\eta_{\rm saddle} = 2X^2 V''/3V$, with
the primes denoting differentiation with respect to $Y$ and with
the result evaluated at the saddle point: $Y = 0$ and $X = X_{\rm
saddle}$. This leads to the following slow-roll parameters
\begin{equation}
    \epsilon_{\rm saddle}=0, \qquad \quad \eta_{\rm saddle}= -0.006097 \,,
\end{equation}
for the values of the parameters taken above. It should be
emphasized that successful inflation is not generic and requires a
tuning of parameters at the level of a part in 100.

Such an adjustment need not be a problem for this model because it
can predict eternal inflation. If causally separated regions of
the universe randomly sample different vacua, then even if the
probability for inflation in any one domain may be strongly
suppressed, the possibility of having eternal inflation infinitely
rewards those domains where inflation occurs (see \cite{LLM} for a
discussion of this issue). It should be stressed that eternal
inflation is not an automatic property of any old inflationary
model. But inflation is eternal in all models where it occurs near
the flat top of an effective potential, with a nearby fall-off to
local minima (such as arises here)
\cite{Eternal,andreivilenkin}.\footnote{Consequently this
particular model does not suffer from the `overshoot' problem of
ref.~\cite{BdA}.}

The spectrum of CMB fluctuations predicted by this scenario is
worked out in detail in ref.~\cite{racetrackinflation}. The
spectrum is typically red, as is typical for inflationary models
having negative curvature, with a representative parameter set
giving $n_s\cong 0.95$ in the COBE region of the spectrum. For the
same parameter choices, the slope of the spectrum is $dn_s/d\ln
k\cong -0.001$, which is negligible in comparison with the current
experimental sensitivity. It is interesting that observations will
eventually be able to discriminate for or against this kind of
model in the relatively near future. Measurable tensor
fluctuations are not expected in this model since the scale of
inflation is $V^{1/4} \sim 10^{14}$ GeV, which is well below the
$3\times 10^{16}$ GeV threshold needed for producing observable
gravity waves.

A particularly interesting feature of the potential generated by
the superpotential, eq.~(\ref{modracetrack}), is that it provides
what is perhaps the simplest realization of what has recently come
to be called the string-theory landscape \cite{landscape}. This
landscape consists of a vast number of local minima having few or
no moduli, which can reasonably be expected to arise for any
system having very many degrees of freedom (like string theory)
once the protection afforded to moduli is removed by supersymmetry
breaking. In particular, the modified racetrack superpotential can
have an enormous number of physically distinct vacua if either of
the parameters $a$ or $b$ should be irrational. A great virtue of
the supergravity obtained using this superpotential is the ability
to explicitly explore these vacua, and so to see what physical
implications their existence may imply. For instance,
ref.~\cite{racetrackinflation} provides a preliminary examination
of the implication of such an enormous number of distinct vacua
for the cosmological constant problem.

\section{Brane-Antibrane Inflation}

The main idea identified to date for obtaining inflation from
string theory in a calculable way is based on the relative motion
of branes and anti-branes through a background geometry under the
influence of their mutual attraction. In this picture it is the
relative motion of the branes and anti-branes which plays the role
of the inflaton, rather than a geometrical mode as in the
Racetrack Inflation discussed in the previous section.

The first proposal to use the relative motion of a configuration
of branes as an inflaton \cite{DvaliTye} supposed these branes all
to be BPS, {\it i.e.} that the branes all preserved a common
supersymmetry and so experienced no inter-brane forces in the
static limit. Although it was promising that the resulting
inter-brane potential should therefore by necessity be very
shallow, the fact that the corrections to exact flatness must be
nonperturbative proved to be an obstacle to further progress
through explicit calculations.

The possibility of calculable inflaton potentials emerged with the
suggestion to use the relative motion of both branes and
brane-antibranes as the inflaton, as was first simultaneously
proposed in refs.~\cite{braneantibraneinflation,dvalishafi}. The
calculability of this scenario is founded on the long-range forces
which branes and anti-branes experience due to the mediation of
massless fields like gravity in the intervening bulk. At first
sight these forces might be expected to be sufficiently weak at
large distances to produce slow-roll inflation, due to the
$r^{-k}$ falloff of the Coulomb interaction in $d_\perp = k + 2$
transverse dimensions \cite{dvalishafi}. However on closer
inspection this expectation proves to be incorrect within a
compact space, due to the inability to separate branes by more
than the size of the internal dimensions within which they live
\cite{braneantibraneinflation}.

The problem with compact spaces can be seen directly from the
predictions for the slow-roll parameters when $V(r) = A + B/r^k$
\cite{braneantibraneinflation}:
\begin{equation}
    \eta = \frac{M_p^2 \, V''}{V} \approx \frac{ck(k+1)B}{A} \, \left(
    \frac{R}{r} \right)^{k+2} \,,
\end{equation}
where it is assumed that the constants $A$, $B$ and $c$ are set
(up to order-unity factors) by the appropriate power of the string
length, $\ell_s$, and that the inter-brane separation, $r$, is
much larger than $\ell_s$. In this expression the constant, $c$,
and the volume, $V_\perp = R^{k+2}$, of the transverse extra
dimensions enters through its appearance in the 4D Planck mass,
$M_p^2 = c V_\perp$. Given that the constants $A$, $B$ and $c$ are
set by the string scale, it is clear that $\eta$ generically
cannot be small because the inter-brane separation typically
cannot exceed the size of the space, $r \le R$.

Because of this last observation the search for inflation becomes
a search for special configurations for which inter-brane forces
are particularly weak, and there have been a number of proposals
for doing so. The first suggestion involved minimizing the net
inter-brane force by placing the branes and anti-branes at
antipodal points in the compact space
\cite{braneantibraneinflation}, but this runs into difficulties
due to the necessity for projecting out bulk-space zero-modes when
computing the inter-brane interactions \cite{kklmmt} (see also
\cite{jellium}). Other proposals include using branes which are
slightly misaligned relative to one another ({\it i.e.} branes at
angles) \cite{angles}, using bulk moduli as the inflaton but with
inter-brane interactions as the driving force \cite{bi2}, using
the $N=2$ supersymmetry of D3-D7 brane interactions \cite{d3d7},
and so on.

The approach within which the modulus-stabilization issue is best
addressed (so far) uses a minor extension \cite{kklmmt} of the
stabilization construction of ref.~\cite{KKLT}. In this
construction the underlying warped Calabi-Yau space is chosen with
its fluxes adjusted to ensure the presence a long warped throat,
whose tip consists of a smoothed-off singularity. Within this
throat the geometry of 5 of the dimensions
--- consisting of the large 4D spacetime directions together with
the radial coordinate which measures the distance down the throat
--- is approximately described by 5D anti-de Sitter space.
Following \cite{KKLT}, one or more anti-D3 branes is imagined to
reside at the tip of this throat, and this breaks the 4D $N=1$
supersymmetry of the background geometry. Finally, a D3 brane is
imagined to be sliding down the throat towards the anti-brane, and
it is the relative motion of these last-mentioned branes which
plays the role of the inflaton.

Inflation may be obtained in this picture, but only at the expense
of carefully adjusting the system parameters to the (0.1 - 1)\%
level. The difficulty with obtaining inflation arises because of
an interplay between the potential which stabilizes the moduli and
the potential of the putative inflaton \cite{kklmmt}. From the
point of view of the low-energy supergravity, this is a special
case of the well-known $\eta$-problem \cite{etaproblem} with
obtaining inflation from a supergravity $F$-term potential.

\begin{figure}[htbp]
\epsfxsize=8cm \centerline{\epsfbox{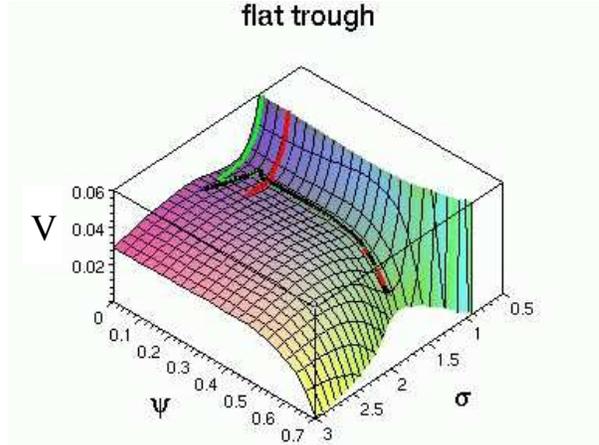}} \caption{Plot
showing the inflationary potential as a function of the
inter-brane separation, $\psi$, and the internal-space volume,
$\sigma$. Also shown are several inflaton trajectories starting
from different initial field values. Units are $M_p=1$.}
\label{F3}
\end{figure}

A further generalization of this construction also allows a
semi-realistic low-energy world to be located within this
scenario, by locating the Standard Model onto a system of
intersecting D3 and D7 branes somewhere within the 6 internal
dimensions \cite{realinf}. Identifying where potentially realistic
low-energy physics can reside inside this kind of inflationary
construction can be important, inasmuch as this opens up the
possibility of asking key post-inflationary questions, like the
nature of the reheating which ultimately leads to our later Hot
Big Bang. Calculations of the density-fluctuation spectrum to be
expected in these models differs from the racetrack models
discussed earlier inasmuch as the spectral index tends to be
slightly bluer, with $n_s \sim 1.03 - 1.08$ \cite{realinf}. A plot
of several successful inflaton trajectories superimposed onto the
effective inflaton potential is shown in Fig.~\ref{F3}.

The possibility of having inflation occur within a throat having
strong warping at its tip might open up several attractive
possibilities for cosmology. In particular, since the warping
tends to reduce the effective tension of the D3 brane as it falls
down the throat, the energy density released by the final
brane/anti-brane annihilation is likely to be set by a lower scale
than the string scale. If this sufficiently suppresses the energy
density released when the brane and anti-brane annihilate compared
with the string scale, it can suppress the fraction of energy
which ends up in the bulk relative to the energy which gets dumped
into spectator branes. If so, then the warping of the throat may
act to improve the efficiency with which inflationary energy gets
converted into reheating standard model degrees of freedom as
opposed to populating phenomenologically problematic bulk states.

An exciting possibility which has emerged from these studies is
the possibility of generating cosmic strings at the end of
inflation \cite{braneantibraneinflation,cosmicstrings}. Since
these need not arise within the racetrack inflation scenario
discussed in the previous section, the formation of such strings
is not a generic prediction of string theory and inflation. But
such strings are generic to brane/anti-brane models, and so long
as the resulting strings are sufficiently long-lived and have a
reasonable energy density they could be detectable through their
effects on CMB fluctuations.

\section{Discussion}

The search for inflation within string theory is still in its
infancy, but considerable progress has been made over recent years
due to the incorporation of the existence of branes into the
search. Branes help with the search for inflation because they can
source flux fields which stabilize moduli, and because their slow
relative motion followed by their annihilation can provide an
attractive geometric realization of hybrid inflation
\cite{hybrid}.

To date two pictures have emerged as to which field might play the
role of the inflaton: either inter-brane separations (brane
inflation) or geometric moduli (racetrack inflation). Explicit
models exist which lead to density fluctuations for which $n_s -
1$ has either sign, and it is intriguing that of the
presently-known models racetrack inflation seems to give $n_s < 1$
while brane-inflation models tend to give $n_s > 1$. Whether this
correlation is robust or is just an artefact of the first few
models is not yet clear. For some of the constructions presently
known it is also possible to identify where our low-energy world
resides, making possible more detailed discussions of reheating
and defect formation.

Although it is possible to obtain 60 $e$-foldings of inflation in
these models, this is not generic and having this much inflation
requires some fine-tuned adjustment of the vacuum properties. One
point of view to take is that in a universe for which different
vacua are chosen in different causal regions, those rare domains
which inflate become infinitely rewarded by dominating the volume
of the subsequent universe. This is a particularly attractive
point of view in the presence of eternal inflation, such as
generically arises when inflation occurs near the local maximum of
a potential with multiple nearby local minima. This is the
situation encountered in the racetrack inflation solutions
discussed in \S3.

It is more difficult to take such a sanguine attitude for the
brane inflation scenarios which have been found up to this point,
such as those of \S4. This is because these fall into the hybrid
inflation category, which need not produce eternal inflation. In
this case it is worth standing back and asking whether string
theory is trying to tell us something if inflation were to arise
in this way, and if it remains true that inflation is not so easy
to achieve.

Although it is comparatively difficult to produce a full 60
$e$-foldings within a hybrid inflation scenario like
brane/anti-brane inflation, 10 to 20 $e$-foldings are much easier
to obtain. The problem basically arises because the scale of the
potential in these models is set by the string scale, and there
are no small numbers to use to make the inflaton potential
sufficiently shallow. Perhaps this suggests that string theory
prefers only to give a small number of $e$-foldings during any
inflationary phase up at the string scale (which is where the
observed temperature fluctuations in the microwave background are
produced). If so, any remaining $e$-foldings of inflation required
by the flatness and homogeneity problems would have to occur
during later epochs, perhaps at the weak scale due to the
super-symmetry breaking potential for some of the ubiquitous
supersymmetric moduli. This suggests a picture of two-stage, late
time inflation \cite{twostep}. If so, it would provide a natural
explanation for effects like a running spectral index, which may
have been observed in the primordial fluctuation spectrum. More
detailed studies of cosmologies of this sort are warranted given
the motivation this kind of picture may receive both from string
theory and the current data.

In summary, inflation is beginning to be found within effective
field theories which capture the essential features of the
low-energy limit of some string vacua having most of their moduli
fixed at the string scale. But this is likely just the beginning
of the exploration of inflation in string vacua, and much remains
to be done towards the goal of a systematic searching for
string-based inflation.

\section*{Acknowledgements}

My part of the research described here has been performed together
with a number of friends and collaborators: J.J.~Blanco-Pillado,
J.M.~Cline, C.~Escoda, M.~G\'omez-Reino, R.~Kallosh, A.~Linde,
M.~Majumdar, the late D.~Nolte, F.~Quevedo, G.~Rajesh and
R.~J.~Zhang. Funding has come from grants from NSERC (Canada),
NATEQ (Qu\'ebec) and McGill University.


\begin{thebibliography}{99}

\bibitem{inflation}
A.H.~Guth (SLAC), Phys.\ Rev.\ {\bf D23} (1981) 347--356;
%
A.~Albrecht, P.J.~Steinhardt, Phys.\ Rev.\ Lett. {\bf 48} (1982)
1220--1223;
%
A.D. Linde, Phys.\ Lett.\ {\bf B108} (1982) 389--393.

\bibitem{cmbinflation}
H.V. Peiris, {\it et. al.}, [astro-ph/0302225];
%
V. Barger, H.-S. Lee and D. Marfatia, [hep-ph/0302150];
%
B. Kyae and Q. Shafi, [astro-ph/0302504];
%
J.R. Ellis, M. Raidal and T. Yanagida, [hep-ph/0303242];
%
A.~Lue, G.D.~Starkman and T.~Vachaspati, [astro-ph/0303268];
%
S.M Leach and A.R.~Liddle [astro-ph/0306305].

\bibitem{L&L}
A.~Liddle and D. Lyth, ``Cosmological Inflation and Large Scale
Structure'', Cambridge University Press (2000).

\bibitem{stringinflationreviews}
For reviews with references, see:
%
F.~Quevedo, Class.\ Quant.\ Grav.\  {\bf 19} (2002) 5721,
[hep-th/0210292];
%
A.~Linde, ``Prospects of inflation,'' [hep-th/0402051].

\bibitem{moduliproblems}
G.~D.~Coughlan, W.~Fischler, E.~W.~Kolb, S.~Raby and G.~G.~Ross,
``Cosmological Problems For The Polonyi Potential,'' Phys.\ Lett.\
B {\bf 131} (1983) 59;
T.~Banks, D.~B.~Kaplan and A.~E.~Nelson, ``Cosmological
implications of dynamical supersymmetry breaking,'' Phys.\ Rev.\ D
{\bf 49} (1994) 779 [hep-ph/9308292];
B.~de Carlos, J.~A.~Casas, F.~Quevedo and E.~Roulet, ``Model
independent properties and cosmological implications of the
dilaton and moduli sectors of 4-d strings,'' Phys.\ Lett.\ B {\bf
318} (1993) 447 [hep-ph/9308325].

\bibitem{omissions}
M. Alishahiha, E. Silverstein and D. Tong, ``DBI in the sky,''
(hep-th/0404084);
%
J.~P.~Hsu, R.~Kallosh and S.~Prokushkin, ``On brane inflation with
volume stabilization,'' JCAP {\bf 0312} (2003) 009
[arXiv:hep-th/0311077];
F.~Koyama, Y.~Tachikawa and T.~Watari, ``Supergravity analysis of
hybrid inflation model from D3-D7 system'',
[arXiv:hep-th/0311191];
H.~Firouzjahi and S.~H.~H.~Tye, ``Closer towards inflation in
string theory,'' Phys.\ Lett.\ B {\bf 584} (2004) 147
[arXiv:hep-th/0312020];
J.~P.~Hsu and R.~Kallosh, ``Volume stabilization and the origin of
the inflaton shift symmetry in string theory,'' JHEP {\bf 0404}
(2004) 042 [arXiv:hep-th/0402047];
%
 O.~DeWolfe, S.~Kachru and H.~Verlinde,
``The giant inflaton,'' JHEP {\bf 0405} (2004) 017
[hep-th/0403123].
%
N.~Iizuka and S.~P.~Trivedi, ``An inflationary model in string
theory,'' hep-th/0403203.

\bibitem{GKP}
S.~B.~Giddings, S.~Kachru and J.~Polchinski, ``Hierarchies from
fluxes in string compactifications,'' Phys. Rev. {\bf D66}, 106006
(2002).

\bibitem{Sethi1}
S.~Sethi, C.~Vafa and E.~Witten, ``Constraints on low-dimensional
string compactifications,'' Nucl.\ Phys.\ B {\bf 480} (1996) 213
[hep-th/9606122];
%
K.~Dasgupta, G.~Rajesh and S.~Sethi, ``M theory, orientifolds and
G-flux,'' JHEP {\bf 9908} (1999) 023 [hep-th/9908088].

\bibitem{calabiyau}
P. Candelas, G.T. Horowitz, A. Strominger and E. Witten, Nucl.\
Phys.\ {\bf B258} (1985) 46--74.

\bibitem{KKLT} S.~Kachru, R.~Kallosh, A.~Linde and S.~P.~Trivedi, ``De
Sitter vacua in string theory,'' Phys.\ Rev.\ D {\bf 68} (2003)
046005 [hep-th/0301240].

\bibitem{cremmeretal}
E.~Cremmer, B.~Julia, J.~Scherk, S.~Ferrara, L.~Girardello and
P.~van Nieuwenhuizen, ``Spontaneous Symmetry Breaking And Higgs
Effect In Supergravity Without Cosmological Constant,'' Nucl.\
Phys.\ B {\bf 147}, 105 (1979).

\bibitem{cyreduction}
E.~Witten, ``Dimensional Reduction Of Superstring Models,'' Phys.\
Lett.\ B {\bf 155} (1985) 151;
%
C.~P.~Burgess, A.~Font and F.~Quevedo, ``Low-Energy Effective
Action For The Superstring,'' Nucl.\ Phys.\ B {\bf 272} (1986)
661.

\bibitem{noscale}
E. Cremmer, S. Ferrara, C. Kounnas and D.V. Nanonpoulos,
``Naturally vanishing cosmological constant in $N=1$
supergravity,'' Phys. Lett. {\bf B133}, 61 (1983);
%
J. Ellis, A.B. Lahanas, D.V. Nanopoulos and K. Tamvakis,
``No-scale Supersymmetric Standard Model,'' Phys. Lett. {\bf
B134}, 429 (1984).

\bibitem{gc}
J.~P.~Derendinger, L.~E.~Ibanez and H.~P.~Nilles, ``On The
Low-Energy D = 4, N=1 Supergravity Theory Extracted From The D =
10, N=1 Superstring,'' Phys.\ Lett.\ B {\bf 155} (1985) 65;
%
M.~Dine, R.~Rohm, N.~Seiberg and E.~Witten, ``Gluino Condensation
In Superstring Models,'' Phys.\ Lett.\ B {\bf 156} (1985) 55.

\bibitem{ourgc}
C.P. Burgess, J.-P. Derendinger, F. Quevedo and M. Quir\'os,
``Gaugino Condensates and Chiral-Linear Duality: An
Effective-Lagrangian Analysis'', Phys.\ Lett.\ B {\bf 348} (1995)
428--442, (hep-th/9501065);
%
``On Gaugino Condensation with Field-Dependent Gauge Couplings'',
Ann.\ Phys.\ {\bf 250} (1996) 193-233, (hep-th/9505171).
%

\bibitem{saltatory}
C.~Escoda, M.~G\'omez-Reino and F.~Quevedo, ``Saltatory de Sitter
string vacua,'' JHEP {\bf 0311} (2003) 065 [hep-th/0307160].

\bibitem{bkq}
C.~P.~Burgess, R.~Kallosh and F.~Quevedo, ``de Sitter string vacua
from supersymmetric D-terms,'' JHEP {\bf 0310} (2003) 056
[arXiv:hep-th/0309187].

\bibitem{Eva}
A. Saltman and E. Silverstein, ``The scaling of the no scale
potential and de Sitter model building,'' hep-th/0402135.


\bibitem{Douglas}
F.~Denef, M.~R.~Douglas and B.~Florea, ``Building a better
racetrack,'' JHEP {\bf 0406} (2004) 034 [hep-th/0404257].

\bibitem{Sethi2}
D.~Robbins and S.~Sethi, ``A barren landscape,'' hep-th/0405011.

\bibitem{racetrackinflation}
J.J.~ Blanco-Pillado, C.P.~Burgess, J.M.~Cline, C.~Escoda,
M.~G\'omez-Reino, R.~Kallosh, A.~Linde, and F.~Quevedo,
``Racetrack Inflation,'' (hep-th/0406230).
%

\bibitem{gcproduct}
%
C.P. Burgess, A. de la Macorra, I. Maksymyk and F. Quevedo,
``Supersymmetric Models with Product Gauge Groups and Field
Dependent Gauge Couplings,'' JHEP 9809 (1998) 007 (30 pages)
(hep-th/9808087).
%

\bibitem{racetrack}
N.V. Krasnikov, ``On Supersymmetry Breaking In Superstring
Theories,'' PLB 193 (1987) 37;
%
L.J. Dixon, "Supersymmetry Breaking in String Theory", in The Rice
Meeting: Proceedings, B. Bonner and H. Miettinen eds., World
Scientific (Singapore) 1990;
%
T.R. Taylor, "Dilaton, Gaugino Condensation and Supersymmetry
Breaking", PLB252 (1990) 59
%
B.~de Carlos, J.~A.~Casas and C.~Mu\~noz, ``Supersymmetry breaking
and determination of the unification gauge coupling constant in
string theories,'' Nucl.\ Phys.\ B {\bf 399} (1993) 623
[hep-th/9204012].

\bibitem{LLM}
A.~D.~Linde, ``Initial Conditions For Inflation,'' Phys.\ Lett.\ B
{\bf 162} (1985) 281;
A.~D.~Linde, D.~A.~Linde and A.~Mezhlumian, ``From the Big Bang
theory to the theory of a stationary universe,'' Phys.\ Rev.\ D
{\bf 49}, 1783 (1994) [gr-qc/9306035].

\bibitem{Eternal}
P.~J.~Steinhardt, ``Natural Inflation,'' In: {\it  The Very Early
Universe}, ed. G.W. Gibbons, S.W. Hawking and S.Siklos, Cambridge
University Press, (1983);
%
A.~D.~Linde, ``Nonsingular Regenerating Inflationary Universe,''
Cambridge University preprint Print-82-0554 (1982);
%
A.~Vilenkin, ``The Birth Of Inflationary Universes,'' Phys.\ Rev.\
D {\bf 27}, 2848 (1983).

\bibitem{andreivilenkin}
A.~D.~Linde, ``Monopoles as big as a universe,'' Phys.\ Lett.\ B
{\bf 327}, 208 (1994) [astro-ph/9402031];
%
A.~D.~Linde and D.~A.~Linde, ``Topological defects as seeds for
eternal inflation,'' Phys.\ Rev.\ D {\bf 50}, 2456 (1994)
[hep-th/9402115];
%
A.~Vilenkin, ``Topological inflation,'' Phys.\ Rev.\ Lett.\  {\bf
72}, 3137 (1994).

\bibitem{BdA}
R.~Brustein and S.~P.~de Alwis,
 ``Moduli potentials in string compactifications with fluxes:
 Mapping the discretuum,''
arXiv:hep-th/0402088.

\bibitem{landscape}
R.~Bousso and J.~Polchinski, ``Quantization of four-form fluxes
and dynamical neutralization of the  cosmological constant,'' JHEP
{\bf 0006} (2000) 006;
%
L.~Susskind, ``The anthropic landscape of string theory,''
(hep-th/0302219);
%
T. Banks, M. Dine and E. Gorbatov, ``Is there a string theory
landscape?,'' (hep-th/0309170).

\bibitem{DvaliTye}
G.~R.~Dvali and S.~H.~H.~Tye, ``Brane inflation,'' Phys.\ Lett.\ B
{\bf 450} (1999) 72 (hep-ph/9812483).

\bibitem{braneantibraneinflation}
C.~P.~Burgess, M.~Majumdar, D.~Nolte, F.~Quevedo, G.~Rajesh and
R.~J.~Zhang, ``The inflationary brane-antibrane universe,'' JHEP
{\bf 0107} (2001) 047 (hep-th/0105204).

\bibitem{dvalishafi}
G.~R.~Dvali, Q.~Shafi and S.~Solganik, ``D-brane inflation,''
[hep-th/0105203].

\bibitem{kklmmt}
S.~Kachru, R.~Kallosh, A.~Linde, J.~Maldacena, L.~McAllister and
S.~P.~Trivedi, ``Towards inflation in string theory,'' JCAP {\bf
0310} (2003) 013 (hep-th/0308055).

\bibitem{jellium}
S.~Buchan, B.~Shlaer, H.~Stoica and S.~H.~H.~Tye, ``Inter-brane
interactions in compact spaces and brane inflation,''
hep-th/0311207.

\bibitem{angles}
J.~Garcia-Bellido, R.~Rabadan and F.~Zamora, ``Inflationary
scenarios from branes at angles,'' JHEP {\bf 0201}, 036 (2002);
%
N.~Jones, H.~Stoica and S.~H.~H.~Tye, ``Brane interaction as the
origin of inflation,'' JHEP {\bf 0207}, 051 (2002);
%
M.~Gomez-Reino and I.~Zavala, ``Recombination of intersecting
D-branes and cosmological inflation,'' JHEP {\bf 0209}, 020
(2002).

\bibitem{bi2}
C.~P.~Burgess, P.~Martineau, F.~Quevedo, G.~Rajesh and
R.~J.~Zhang, ``Brane antibrane inflation in orbifold and
orientifold models,'' JHEP {\bf 0203} (2002) 052 [hep-th/0111025].

\bibitem{d3d7}
C.~Herdeiro, S.~Hirano and R.~Kallosh, ``String theory and hybrid
inflation / acceleration,'' JHEP {\bf 0112} (2001) 027
[hep-th/0110271];
%
K.~Dasgupta, C.~Herdeiro, S.~Hirano and R.~Kallosh, ``D3/D7
inflationary model and M-theory,'' Phys.\ Rev.\ D {\bf 65} (2002)
126002 [hep-th/0203019].

\bibitem{etaproblem}
See for instance: E.~J.~Copeland, A.~R.~Liddle, D.~H.~Lyth,
E.~D.~Stewart and D.~Wands, ``False vacuum inflation with Einstein
gravity,'' Phys.\ Rev.\ D {\bf 49} (1994) 6410 [astro-ph/9401011].

\bibitem{realinf}
C.~P.~Burgess, J.~M.~Cline, H.~Stoica and F.~Quevedo, ``Inflation
in realistic D-brane models,'' [hep-th/0403119].

\bibitem{cosmicstrings}
S.~Sarangi and S.~H.~H.~Tye, ``Cosmic string production towards
the end of brane inflation,'' Phys.\ Lett.\ B {\bf 536} (2002) 185
[hep-th/0204074];
%
G.~Dvali, R.~Kallosh and A.~Van Proeyen, ``D-term strings,'' JHEP
{\bf 0401} (2004) 035 [hep-th/0312005];
G.~Dvali and A.~Vilenkin, ``Formation and evolution of cosmic
D-strings,'' JCAP {\bf 0403} (2004) 010 [hep-th/0312007];
E.~J.~Copeland, R.~C.~Myers and J.~Polchinski, ``Cosmic F- and
D-strings,'' JHEP {\bf 0406} (2004) 013 [hep-th/0312067];
L.~Leblond and S.~H.~H.~Tye, ``Stability of D1-strings inside a
D3-brane,'' JHEP {\bf 0403} (2004) 055 [hep-th/0402072];
K.~Dasgupta, J.~P.~Hsu, R.~Kallosh, A.~Linde and M.~Zagermann,
``D3/D7 brane inflation and semilocal strings,'' (hep-th/0405247).

\bibitem{hybrid}
A.~D.~Linde, ``Hybrid inflation,'' Phys.\ Rev.\ D {\bf 49} (1994)
748 [astro-ph/9307002].

\bibitem{twostep}
D.~H.~Lyth and E.~D.~Stewart, ``Thermal inflation and the moduli
problem,'' Phys.\ Rev.\ D {\bf 53}, 1784 (1996) [hep-ph/9510204];
J.~A.~Adams, G.~G.~Ross and S.~Sarkar, ``Multiple inflation,''
Nucl.\ Phys.\ B {\bf 503} (1997) 405 [hep-ph/9704286];
G.~German, G.~G.~Ross and S.~Sarkar, ``Implementing quadratic
supergravity inflation,'' Phys.\ Lett.\ B {\bf 469} (1999) 46
[hep-ph/9908380];
``Low-scale inflation,'' Nucl.\ Phys.\ B {\bf 608} (2001) 423
[hep-ph/0103243].

\end{thebibliography}
\end{document}